\documentstyle[epsfig,latexsym,amssymb,aps,floats,eqsecnum,preprint,
amsfonts]{revtex}
\input epsfig.sty
\def\laq{\raise 0.4ex\hbox{$<$}\kern -0.8em\lower 0.62 ex\hbox{$\sim$}}
\def\gaq{\raise 0.4ex\hbox{$>$}\kern -0.7em\lower 0.62 ex\hbox{$\sim$}}

\begin{document}
\begin{titlepage}
\begin{flushright}
CERN-TH/2003-231
\end{flushright}
\vspace*{1cm}

\begin{center}
{\large{\bf Tracking curvaton(s)?}}
\vskip 1 cm 
{\sl Massimo Giovannini\footnote{Electronic 
address: massimo.giovannini@cern.ch}}
\vskip 0.5 cm 
{\sl  Theoretical Physics Division, CERN, CH-1211 Geneva 23, 
Switzerland}
\vspace{1cm}
\noindent
\begin{abstract}
The ratio of the curvaton energy density to that of the dominant 
component of the background sources may be constant during a significant period in 
 the evolution of the Universe. 
The possibility of having tracking curvatons, whose decay occurs prior to the nucleosynthesis 
epoch, is studied. It is argued that the tracking curvaton dynamics is 
disfavoured since the value of the curvature perturbations prior to curvaton decay 
is smaller than the value required by observations. It is 
also argued, in a related context, that the minimal inflationary curvature scale 
compatible with the curvaton paradigm may be lowered in the case of low-scale quintessential 
inflation.
\end{abstract} 
\vskip0.5pc
\end{center}
\end{titlepage}
\newpage
\noindent

\renewcommand{\theequation}{1.\arabic{equation}}
\setcounter{equation}{0}
\section{Introduction} 
The isocurvature fluctuations generated by a field $\epsilon$ 
that is 
light during the inflationary phase and  later on decays, 
can efficiently produce adiabatic curvature perturbations.  
This idea has recently been discussed  in different,  frameworks 
ranging from the context of conventional inflationary models \cite{lyth1,moroi}
to the case of pre-big bang models\cite{enqvist,giovannini1,sloth}. It is 
appropriate to recall that an earlier version of this proposal 
was discussed in \cite{mollerach}, not with the specific purpose of converting 
isocurvature into adiabatic modes, but rather with the hope 
of providing physical initial conditions for the baryon isocurvature 
perturbations.

A common feature 
of various scenarios 
\cite{bartolo,lyth2,giovannini2,basterogil,enqvist1,enqvist2,lyth3,lyth4,dimopoulos,fujii} 
is that the energy density of the 
homogeneous component of this light curvaton field decreases more
slowly than the energy density of the background geometry.
In the simplest realization, the Universe  
suddenly becomes dominated by radiation, as soon as inflation ends. 
During the radiation epoch, the homogeneous component 
of the curvaton field is roughly constant down to the moment 
when the Hubble parameter 
is comparable with the curvaton mass, i.e.  $H \sim m$.
The energy 
density of the oscillating curvaton field 
 then decreases as $a^{-3}$, where $a$ is the 
scale factor of a conformally flat Friedmann--Robertson--Walker 
(FRW) Universe. Since the energy density of the background
radiation scales as $a^{-4}$, the ratio 
\begin{equation}
r(a) = \frac{\rho_{\epsilon}(a)}{\rho_{\rm r}(a) } 
\label{r}
\end{equation}
will increase as $a$. Both a different post-inflationary 
history and a different potential will produce deviations from this 
scaling law. In particular, if the potential is not exactly quadratic 
(or if the coupling of the curvaton to the geometry is not 
exactly minimal) \cite{lyth3}
$r(a)$ may decrease. If the post-inflationary phase 
is not immediately dominated by radiation, $r(a)$ may increase 
even faster than $a$. If, after the end of inflation the 
inflaton field turns into a quintessence-like field \cite{pv} 
(see also \cite{dimopoulos,giovannini3}), the background 
energy density will be dominated by the kinetic term of the 
quintessence field. In this case $r(\eta)$ increases 
as $a^3$ \cite{giovannini2}. 

It is not impossible
that the ratio $r(a)$ stays constant during a 
significant period in the post-inflationary evolution.
This subject  will be explored in the 
present investigation. It will be argued 
that this type of ``tracking'' phase is disfavoured because 
the initial isocurvature mode will (rather generically) turn
into an adiabatic mode of much smaller amplitude,
making the whole scenario uneffective in correctly 
reproducing the amplitude of the large-scale fluctuations. 

One of the reasons to speculate on the possibility 
of tracking curvatons is that one would like to allow 
for models where the curvature scale, at 
the end of inflation, is much smaller than the value required, for 
instance, in single field inflationary models, where the 
curvature fluctuations are directly amplified during the 
inflationary phase. It  will be shown that if the inflationary 
phase is not followed suddenly by radiation but rather by a kinetic 
phase (as in the case of quintessential inflation) the minimal 
allowed curvature scale is a bit smaller than in the standard case 
of sudden radiation domination.

The present article is organized as follows. In Section
II the basic problem will be formulated. In Section III
it will be shown that if the curvaton field tracks  the evolution 
of the dominant component of the background the 
resulting adiabatic mode is 
smaller then the initial isocurvature mode.
In Section IV the lower bound on the inflationary 
curvature scale will be discussed. Finally, Section V contains 
some concluding remarks.

\renewcommand{\theequation}{2.\arabic{equation}}
\setcounter{equation}{0}
\section{Formulation of the problem} 

For illustrative purposes it is useful to 
consider the case where the inflationary phase is suddenly replaced 
by a radiation-dominated phase.
During inflation, the curvaton field $\epsilon$ should be nearly massless and 
displaced from the minimum of its potential $W(\epsilon)$:
\begin{equation}
\frac{\partial^2 W}{\partial\epsilon^2} \ll H_{\rm i}^2,~~~~~~~~~|\epsilon_{\rm i} -\epsilon_{0}| > H_{\rm i},
\label{p2}
\end{equation}
where  the subscript denotes the moment 
when the cosmologically interesting scales left the horizon during inflation. 
Under the assumption that the Universe is dominated by radiation, 
the evolution equations for the background fields can be written\footnote{ Units 
$M_{\rm P} = (8 \pi G)^{-1/2}$ will be adopted. The variable $\eta$ denotes 
the conformal time coordinate while $t$ denotes the cosmic time.Derivatives with 
respect to either conformal or cosmic time will be denoted, respectively, by a prime and by 
an overdot. Finally, the expansion rate, in the conformal and cosmic time coordinates 
will be denoted by ${\cal H} = a'/a$ and by $H = \dot{a}/a$. }
\begin{eqnarray}
&& M_{\rm P}^2 {\cal H}^2 = \frac{a^2}{3} (\rho_{\epsilon} + \rho_{\rm r}),
\label{001}\\
&& M_{\rm P}^2 {\cal H}' = - \frac{1}{3} \biggl[ a^2 \rho_{\rm r} + {\epsilon'}^2 - W a^2\biggr],
\label{ij1}\\
&&  \epsilon'' + 2 {\cal H} \epsilon' + \frac{\partial W}{ \partial \epsilon} a^2 = 0,
\label{eps1}\\
&& \rho_{\rm r}' + 4 {\cal H} \rho_{\rm r} =0,
\label{rho1}
\end{eqnarray}
where it has been assumed that the curvaton is minimally coupled to the background geometry. This assumption may be relaxed by requiring, 
for instance, that the coupling be non-minimal. This would imply the addition, in Eq. (\ref{eps1}), of a term going like 
$c R \epsilon$, where $R\propto H^2$ is the Ricci scalar of a radiation-dominated FRW background. The addition 
of such a term is expected to lead to a decrease of $r(a)$ \cite{lyth3} even prior to the true oscillatory phase  taking place when $H\sim m$. 
For the purposes of the present Section, it is useful
to parametrize  the evolution of $\epsilon$ in terms of $r = \rho_{\epsilon}/\rho_{\rm r}$.
In this case 
\begin{eqnarray}
&& {\epsilon'}^2 = - M_{\rm P}^2 {\cal H} a^{4} \biggl(\frac{r}{a^4}\biggr)',
\nonumber\\
&& \frac{\partial W}{\partial \epsilon} a^2 = - \frac{1}{a^2} ( a^2 \epsilon')',
\label{p3}
\end{eqnarray}
where the first equation has been  obtained by taking the derivative of the definition of $r$ 
and by subsequent use of Eq. ({\ref{eps1}).

Provided $r(\eta_{{\rm i}}) \ll 1$, 
 three possible physical situations may arise. If $r' >0$, the curvaton field 
will become, at some point, dominant over the background density. If $r' <0$ ,the curvaton field 
will never dominate over the background geometry. The third  possibility is to have  $r' \simeq 0$, which is the case
to be discussed in the following. 
Even if examples will be provided in the framework 
of specific potentials, it is better, at this stage, to think of the potential as a function 
of the ratio $r$, as suggested by Eqs. (\ref{p3}).

In order to study the evolution of the fluctuations consider the general 
form of the perturbation equations, namely: 
\begin{eqnarray}
&&\nabla^2 \Phi - 3 {\cal H} ( {\cal H} \Phi + \Phi') = \frac{a^2}{2 M_{\rm P}^2}\rho_{\rm r} \delta_{\rm r} + 
\frac{1}{2 M_{\rm P}^2}\biggl[ - \Phi {\epsilon'}^2 + \epsilon' \chi' + \frac{ \partial W}{\partial \epsilon}a^2 \chi \biggr],
\label{00p}\\
&& \Phi'' + 3 {\cal H} \Phi' + ( {\cal H}^2 + 2 {\cal H}') \Phi = \frac{a^2}{6 M_{\rm P}^2} \rho_{\rm r} \delta_{{\rm r}} - 
\frac{1}{2 M_{\rm P}^2 } \biggl[ \Phi {\epsilon'}^2 - \epsilon' \chi' + \frac{\partial W}{\partial\epsilon} a^2 \chi \biggr],
\label{ijp}\\
&& {\cal H} \Phi +  \Phi' = \frac{\epsilon'}{2 M_{\rm P}^2} \chi + \frac{ 2}{3 M_{\rm P}^2} u a^2 \rho_{\rm r},
\label{0i}\\
&& \chi'' + 2 {\cal H} \chi' - \nabla^2 \chi + \frac{\partial^2 W}{\partial \epsilon^2} a^2 \chi - 4 
\epsilon' \Phi' + 2 \frac{\partial W}{\partial \epsilon} a^2 \Phi =0,
\label{chip}\\
&& \delta_{\rm r}' - 4 \Phi' - \frac{4}{3} \nabla^2 u =0,
\label{deltar}\\
&& u' - \frac{1}{4} \delta_{\rm r} - \phi =0,
\label{u}
\end{eqnarray}
where $\Phi$ corresponds to the  gauge-invariant (longitudinal) fluctuation of the geometry, $\chi$ is the 
curvaton fluctuation; $u$ and $\delta_{\rm r}$ are, respectively, the gauge-invariant velocity field and the gauge-invariant density 
contrast. 
Introducing the useful notation $ x = \ln{(\eta/\eta_{\rm i})}$,
 Eqs. (\ref{00p}), (\ref{ijp}) and (\ref{chip}) 
can be written, in Fourier space,
\begin{eqnarray}
&& \frac{d\Phi_{k}}{d x}  + \Phi_{k} = - 
\frac{\delta_{\rm r}(k)}{2} - \frac{1}{6 M_{\rm P}^2}\biggl[ - \Phi_{k} \biggl(\frac{d \epsilon}{d x}\biggr)^2 +
\frac{d \epsilon}{d x} \frac{d \chi_{k}}{d x} + e^{4 x} \eta_{\rm i}^2 \frac{\partial W}{\partial\epsilon} \chi_{k} \biggr],
\label{00p2}\\
&& \frac{ d^2 \Phi_{k}}{d x^2} + 2  \frac{d \Phi_{k}}{d x} - \Phi_{k} = \frac{\delta_{\rm r}(k)}{2} - 
\frac{1}{2 M_{\rm P}^2} \biggl[ \Phi_{k} \biggl( \frac{d \epsilon}{d x} \biggr)^2 - \frac{d \epsilon}{d x} \frac{d \chi_{k}}{d x} +  
e^{4 x} \eta_{i}^2 \frac{\partial W}{\partial \epsilon} \chi_{k} \biggl],
\label{ijp2}\\
&& \frac{d^2 \chi_{k}}{d x^2} + \frac{ d \chi_{k}}{d x} + e^{4 x} \eta_{\rm i}^2 \frac{\partial^2 W}{\partial \epsilon^2} \chi_{k}  
-4 \frac{d \epsilon}{d x} \frac{ d \Phi_{k}}{d x} + 2 \frac{\partial W}{\partial \epsilon} e^{4 x} \eta_{\rm i}^2 \Phi_{k}=0.
\label{chi2}
\end{eqnarray}
Combining Eqs. (\ref{00p2}) and (\ref{ijp2}), 
\begin{equation}
\frac{ d^2 \Phi_{k}}{ d x^2} + 3 \frac{ d \Phi_{k}}{d x}  + \frac{1}{3 M_{\rm P}^2}
 \biggl[ \Phi \biggl( \frac{d \epsilon}{d x}\biggr)^2 - 
\frac{ d \epsilon}{ d x} \frac{d\chi_{k}}{ d x} + 2 e^{4 x} \eta_{\rm i}^2 \frac{\partial W}{\partial \epsilon} \chi_{k} \biggr] =0.
\label{comb}
\end{equation}
Imposing isocurvature initial conditions right at the onset of the 
radiation dominated epoch implies, for long-wavelength modes:
\begin{equation}
\Phi_{k}(\eta_{\rm i}) =0, ~~~~~\Phi'_{k}(\eta_{\rm i}) =0,~~~~~~\chi_{k}(\eta_{\rm i}) = \chi_{k}^{({\rm i})}, ~~~~~~~
\chi'_{k}(\eta_{\rm i}) =0.
\label{incon1}
\end{equation}
In terms of ${\cal R}$, the gauge-invariant spatial curvature perturbations,
\begin{equation}
{\cal R}_{k} = - \biggl( \Phi_{k} + {\cal H}\frac{{\cal H} \Phi_{k} + \Phi_{k}'}{{\cal H}^2 - {\cal H}'}\biggr),
\end{equation}
 the initial 
conditions given in Eq. (\ref{incon1}) imply ${\cal R}_{k}(\eta_{\rm i}) =0$. From the Hamiltonian 
constraint, recalling Eq. (\ref{incon1}), the relation between the initial density contrast and the 
initial curvaton fluctuation can be obtained 
\begin{equation}
\delta_{\rm r}^{({\rm i})}(k) = -  \biggl[\frac{\partial W}{\partial \epsilon}\biggr]_{\eta_{\rm i}} 
\frac{\chi_{k}^{({\rm i})}}{\rho_{r}^{({\rm i})}},
\label{deltai}
\end{equation}
where $\delta_{\rm r}^{{(\rm i)}}(k) = \delta_{\rm r}(k,\eta_{\rm i})$. 
Using the observation that, from Eq. (\ref{deltar}), the quantity $\delta_{\rm r}(k) 
- 4 \Phi_{k}$ is conserved in the long-wavelength limit the initial value of the isocurvature mode can be 
related to the final values both of the Bardeen potential and of the curvaton fluctuation.
On this basis, useful analytical relations will  be derived and later compared with the numerical solutions.
From  Eq. (\ref{deltai}) and taking into account Eq. (\ref{incon1}), the Hamiltonian constraint 
of Eq. (\ref{00p2}) leads to the relation 
\begin{equation}
\delta_{\rm r}^{({\rm f})}(k) = 4 \Phi_{k}^{({\rm f})} -  \biggl[\frac{\partial W}{\partial \epsilon}\biggr]_{\eta_{\rm i}} 
\frac{\chi_{k}^{({\rm i})}}{\rho_{r}^{({\rm i})}},
\label{deltaf}
\end{equation}
where the superscript ``${\rm f}$'' denotes the final (constant) value of the corresponding quantity.
In the  case $r'\simeq 0$, from  Eq. (\ref{p3})
\begin{equation}
\epsilon' \simeq 2 {\cal H} M_{\rm P} \sqrt{r}.
\end{equation}
Then Eqs. (\ref{00p2})--(\ref{chi2}) admit a solution with constant mode. 
Equation (\ref{chi2}) and the combination of  Eqs. (\ref{00p2}) with (\ref{deltaf}) lead, 
respectively, to the following two relations
\begin{eqnarray}
&& \frac{\partial^2 W}{\partial \epsilon^2} \chi_{k}^{({\rm f})} + 2 \frac{\partial W}{\partial \epsilon } \Phi_{k}^{({\rm f})}=0,
\label{first}\\
&& \Phi_{k}^{({\rm f})} =  \frac{1}{6}  \biggl(\frac{\partial W}{\partial \epsilon}\biggr)_{\eta_{\rm i}} 
\frac{\chi_{k}^{({\rm i})}}{\rho_{r}^{({\rm i})}},
\label{second}
\end{eqnarray}
which allow, in turn, $\chi_{k}^{({\rm f})}$ to be determined in terms of $\chi_{k}^{({\rm i})}$, namely
\begin{equation}
\biggl(\frac{\partial^2 W}{\partial\epsilon^2 }\biggr)_{\eta_{\rm f}}
\chi_{k}^{({\rm f})} = - \frac{1}{3} \biggl(\frac{\partial W}{\partial\epsilon}\biggr)_{\eta_{\rm i}} 
\biggl(\frac{\partial W}{\partial\epsilon}\biggr)_{\eta_{\rm f}} \frac{\chi_{k}^{({\rm i})}}{\rho_{r}^{({\rm i})}}.
\end{equation}
Defining now the constant value of $r$ as $r_{\rm c}$, the previous results imply 
\begin{equation}
 \Phi_{k}^{({\rm f})} \simeq - \frac{\sqrt{r_{\rm c}}}{9}  \tilde{\chi}_{k}^{({\rm i})} + {\cal O}(r_{\rm c}),
\label{first1}
\end{equation}
where $\tilde{\chi}_{k} = \chi_{k}/M_{\rm P}$. From Eq. (\ref{first1})
it also follows, using (\ref{first}) together with (\ref{p3})
\begin{equation}
\chi_{k}^{({\rm f})} \simeq - \frac{2 r_{\rm c}}{9}   \chi_{k}^{({\rm i})}.
\label{second1}
\end{equation}
Eqs. (\ref{first1}) and (\ref{second1}) imply that 
$\Phi_{k}^{({\rm f})}$ is always smaller (by a factor $\sqrt{r_{c}}\ll 1$) 
than the value of the curvaton fluctuations at the end of inflation. Hence, even 
assuming that the curvaton fluctuations are amplified with flat spectrum, i.e. 
$\chi_{k}^{({\rm i})} \sim H_{\rm i} /(2 \pi)$, the final value of the 
produced adiabatic mode  $\Phi_{k}^{({\rm f})}$ will  always be
phenomenologically negligible.  The value of $\chi_{k}^{({\rm f})}$ will be even smaller.
In the second place the above equations are derived assuming that, in the asymptotic regime, 
the following relation holds, 
\begin{equation}
\biggl.\frac{1}{\dot{\epsilon}^2} \biggl(\frac{\partial W}{\partial \epsilon}\biggr)^2 \biggr|_{\eta_{\rm f}} \simeq \frac{1}{4} 
 \biggl.\biggl(\frac{\partial^2 W}{\partial \epsilon^2}\biggr) \biggr|_{\eta_{\rm f}}.
\label{relpot}
\end{equation}
The relation (\ref{relpot}) holds exactly in the case of exponential 
potentials of the type $W(\epsilon)= W_0 e^{- \epsilon/\Lambda}$. In this case 
the solution for $\epsilon$ is obtained by solving Eq. (\ref{eps1}) in a 
radiation-dominated background, with the result that  
\begin{equation}
\epsilon(\eta) = \epsilon_{\rm i} + \epsilon_{1} \ln{(\eta/\eta_{\rm i})},
\end{equation}
where $\epsilon_1 = 4 \Lambda$ and $ r_c = 4 (\Lambda/M_{\rm P})^2$.
This remark already rules out a curvaton potential, which is 
purely exponential down to the moment of curvaton decay. In this case 
Eqs. (\ref{first1}) and(\ref{second1}) imply the smallness of the obtained curvature 
perturbations. However it is also suggestive to think of the 
case where the curvaton potential is not purely exponential \cite{barreiro,albrecht}.
The question would be, in this case, if the new features of the potential 
allow a radically different dynamics of the fluctuations.

\renewcommand{\theequation}{3.\arabic{equation}}
\setcounter{equation}{0}
\section{Explicit examples}
In order to discuss a physically realistic situation,
consider as an example the following potential 
\begin{equation}
W(\epsilon) = W_0 [ \cosh{(\epsilon/\Lambda)} -1]^{\alpha},
\label{class}
\end{equation}
which has been studied, for instance, in the context 
of quintessence models \cite{sahni}. In spite of this formal 
analogy, in the present case, the field $\epsilon$ will decay 
prior to big bang nucleosynthesis and it will not act as a quintessence field. 
With this caveat, also other 
quintessential potentials (see, for instance, \cite{rp1} for a comprehensive review) 
may be used, in the present context, 
to construct physical models.
The class  of potentials (\ref{class})
 has the property that when $|\epsilon/\Lambda| \gg 1$ 
the potential admits  solutions where the ratio between the 
curvaton energy density and the radiation energy density is, momentarily, constant. On the other 
hand, when $ |\epsilon/\Lambda| \geq 1$ the curvaton oscillates.
Consider, for concreteness, the simplest case, namely
\begin{equation}
W(\epsilon) = M^4 [ \cosh{(\epsilon/\Lambda)} -1],
\end{equation}
where the oscillations are quadratic since, for 
$|\epsilon/\Lambda|\ll 1$,
\begin{equation}
W(\epsilon) \simeq \frac{M^4}{2 \Lambda^2} \epsilon^2.
\end{equation}
Using the  notation 
\begin{equation}
\tilde{\epsilon} = \frac{\epsilon}{M_{\rm P}},~~~~~~\mu = \frac{\Lambda}{M_{\rm P}}, ~~~~~
\nu = \frac{M}{M_{\rm P}}, ~~~~~~~~~\xi = \frac{H_{\rm i}}{M_{\rm P}},
\end{equation}
the evolution of the curvaton is determined by the following equation
\begin{equation}
\frac{d^2\tilde{\epsilon} }{dx^2} + \frac{d \tilde{\epsilon}}{d x} + e^{4 x} \frac{\nu^4}{\mu \xi^2} \sinh{(\tilde{\epsilon}/\mu)}=0.
\end{equation}
The constraints of Eq. (\ref{p2}) imply, in the present case, 
\begin{equation}
\frac{M^4}{\Lambda^2} e^{ - \epsilon_{\rm i}/\Lambda} \ll H_{\rm i}^2,~~~~~~~|\epsilon_{\rm i}| > \Lambda, ~~~~~~~ |\epsilon_{\rm i}| \gg H_{\rm i}.
\label{cond2}
\end{equation}

The ratio between the curvaton energy density and the radiation energy density can 
 be expressed as
\begin{equation}
r(x) = \frac{1}{6} \biggl(\frac{ d \tilde{\epsilon}}{d x}\biggr)^2 + \frac{\nu^4}{3 \xi^2} 
e^{4 x} [\cosh{ (\tilde{\epsilon}/\mu)} -1].
\label{rx}
\end{equation}
The initial data (obeying the conditions (\ref{cond2}))
 for the evolution of the homogeneous 
component of the curvaton field are set during the inflationary epoch in such a way that 
$|\tilde{\epsilon}_{\rm i} |> \mu$. Since  the potential is essentially 
exponential, in this regime the field will be swiftly attracted towards the tracking solution, where 
the critical fraction of curvaton's energy density will be constant.
In this regime the solution can be approximated by 
\begin{equation}
\tilde{\epsilon} = - \tilde{\epsilon}_{0} + 4 \mu x, ~~~~~~\tilde{\epsilon}_{0}= \mu
 \ln{\biggl(\frac{ 8 \mu^2 \xi^2}{\nu^4}\biggr)}.
\label{solt}
\end{equation}
When $H \sim M^4/\Lambda^2 \sim M^2 \nu^2/\mu^2$, i.e. 
\begin{equation}
x_{\rm m} = \frac{1}{4}\biggl[ 1 + \ln{\biggl( \frac{8 \mu^2 \xi^2}{\nu^4}\biggr)}\biggr],
\end{equation}
the curvaton will start oscillating 
in the minimum of its potential. During the tracking phase 
$ r (x) \simeq r_{\rm tr} = 4\mu^2 \ll 1$.
During the oscillatory regime the energy density of the curvaton will 
decrease as $a^{-3}$ so that, 
\begin{equation}
r(x) = 4 \mu^2 \biggl(\frac{a}{a_{\rm m}}\biggr), ~~~~~~~~~x \geq x_{\rm m}.
\end{equation} 

Let us now write, as a first step, the evolution equations for the fluctuations 
in the case when the background after the end of inflation is 
suddenly dominated by radiation:
\begin{eqnarray}
&& \frac{d\Phi_{k}}{d x}  + \Phi_{k} = - 
\frac{\delta_{\rm r}(k)}{2} - \frac{1}{6}\biggl[ - \Phi_{k} \biggl(\frac{d \epsilon}{d x}\biggr)^2 +
\frac{d \tilde{\epsilon}}{d x} \frac{d \tilde{\chi}_{k}}{d x} + e^{4 x} \frac{\nu^4}{\mu\xi^2} 
\sinh{(\tilde{\epsilon}/\mu)}
\tilde{\chi}_{k} \biggr],
\label{00p2t}\\
&& \frac{ d^2 \Phi_{k}}{d x^2} + 2  \frac{d \Phi_{k}}{d x} - \Phi_{k} = \frac{\delta_{\rm r}(k)}{2} - 
\frac{1}{2} \biggl[ \Phi_{k} \biggl( \frac{d \tilde{\epsilon}}{d x} \biggr)^2 - \frac{d \tilde{\epsilon}}{d x} 
\frac{d \tilde{\chi}_{k}}{d x} +  
e^{4 x} \frac{\nu^4}{\mu\xi^2} \sinh{(\tilde{\epsilon}/\mu)} \tilde{\chi}_{k} \biggl],
\label{ijp2t}\\
&& \frac{d^2 \tilde{\chi}_{k}}{d x^2} + \frac{ d \tilde{\chi}_{k}}{d x} + e^{4 x}  \frac{\nu^4}{\mu^2 \xi^2} 
\cosh{(\tilde{\epsilon}/\mu)} \tilde{\chi}_{k}  
-4 \frac{d \tilde{\epsilon}}{d x} \frac{ d \Phi_{k}}{d x} + 2 \frac{\nu^4}{\mu\xi^2}e^{4 x} \sinh{(\tilde{\epsilon}/\mu)}\Phi_{k}=0,
\label{chi2t}\\
&& \frac{d}{d x}\biggl[ \delta_{\rm r}(k) - 4 \Phi_{k}\biggr]=0.
\end{eqnarray}
Combining Eqs. (\ref{00p2}) and (\ref{ijp2}) 
\begin{equation}
\frac{ d^2 \Phi_{k}}{ d x^2} + 3 \frac{ d \Phi_{k}}{d x}  + \frac{1}{3}
 \biggl[ \Phi_{k} \biggl( \frac{d \tilde{\epsilon}}{d x}\biggr)^2 - 
\frac{ d \tilde{\epsilon}}{ d x} \frac{d\chi_{k}}{ d x} + 2 e^{4 x} \frac{\nu^4}{\mu\xi^2} \sinh{(\tilde{\epsilon}/\mu)} 
\tilde{\chi}_{k} \biggr] =0.
\label{combt}
\end{equation}
The evolution equations of $\epsilon$ and of the fluctuations can be numerically integrated. Analytical 
approximations can also be obtained. In Fig. \ref{FIG1} the numerical results 
are illustrated, for a typical set of parameters, in terms of $r(x)$ as defined in Eq. (\ref{rx}). It can be 
appreciated that after a phase where $r' \simeq 0$, the energy density of the $\epsilon$ will decrease, implying
that $r(x) \sim e^{x}$. 
\begin{figure}
\centerline{\epsfxsize = 11cm  \epsffile{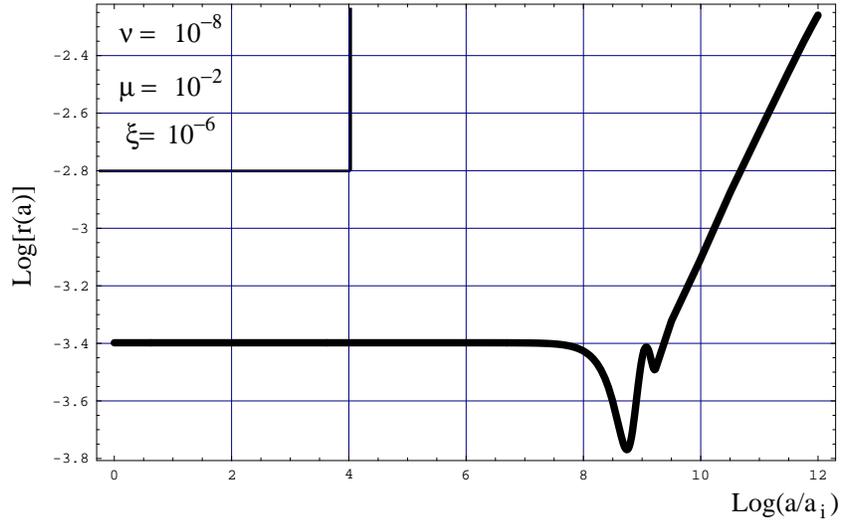}}
\vskip 3mm
\caption[a]{The numerical results for the evolution of $r$ (see Eq. (\ref{rx})) are reported.}
\label{FIG1} 
\end{figure}
In Fig. \ref{FIG2} the evolution of $\epsilon(x)$ is reported. The numerical results (full line) are compared 
with the analytical approximation (dashed line).
\begin{figure}
\centerline{\epsfxsize = 11cm  \epsffile{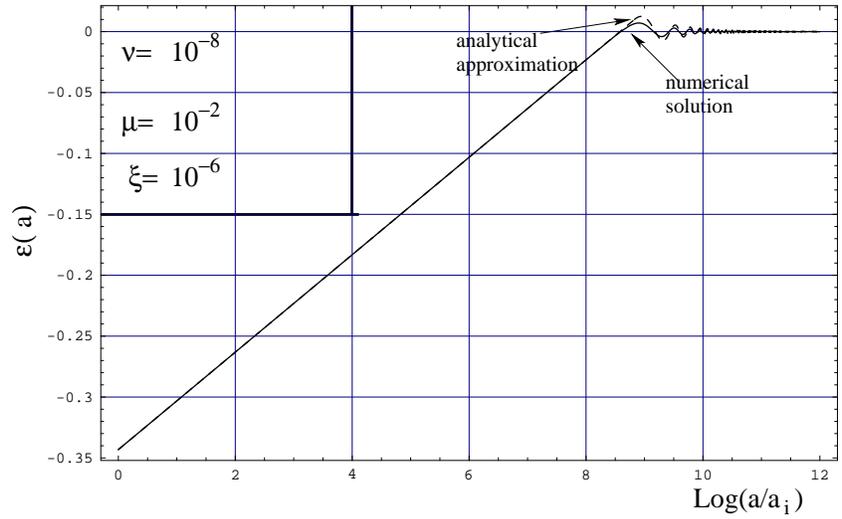}}
\vskip 3mm
\caption[a]{The full line illustrates the numerical result 
while the dashed line shows the analytical approximation based on Eq. (\ref{solosc}).}
\label{FIG2} 
\end{figure}
In fact, a  useful analytical approximation to the whole evolution can be obtained 
by matching the solution (\ref{solt}), valid during the tracking regime, 
with the exact solution (of the approximate potential) valid in the oscillating regime. 
In order to do so it is useful to write the evolution 
equation for $\tilde{\epsilon}$ as 
\begin{equation}
\frac{d^2 \tilde{\epsilon}}{d y^2} + \frac{3}{2 y} \frac{d \tilde{\epsilon}}{d y}  + \tilde{\epsilon}=0,~~~~~~
y = e^{2 x}\frac{\nu^2}{2 \mu \xi},
\end{equation} 
whose solution is 
\begin{equation}
\tilde{\epsilon}(x) = e^{- x/2} \biggl[ A J_{1/4}\biggl(e^{2 x} \frac{\nu^2}{2 \mu\xi}\biggr) 
+ B Y_{1/4}\biggl(e^{2 x} \frac{\nu^2}{2 \mu\xi}\biggr)\biggr], ~~~~~x>x_{\rm m},
\label{solosc}
\end{equation}
where the numerical constants 
\begin{eqnarray}
&& A = \frac{e^{\frac{x_{{\rm m}}}{2}}\,\pi \,\biggl[ -4\,\mu \,\xi \,
       Y_{1/4}(\frac{e^{2\,x_{{\rm m}}}\,{\nu }^2}{2\,\mu \,\xi }) - 
      e^{2\,x_{{\rm m}}}\,{\nu }^2\,Y_{5/4}(\frac{e^{2\,x_{{\rm m}}}\,{\nu }^2}{2\,\mu \,\xi })\,
       \left( 4\,x_{{\rm m}} - \log (\frac{8\,{\mu }^2\,{\xi }^2}{{\nu }^4}) \right)  \biggr] }{4\,\xi },
\nonumber\\
&& B=\frac{e^{\frac{x_{{\rm m}}}{2}}\,\pi \,\biggl[4\,\mu \,\xi \,
       J_{1/4}(\frac{e^{2\,x_{{\rm m}}}\,{\nu }^2}{2\,\mu \,\xi }) + 
      e^{2\,x_{{\rm m}}}\,{\nu }^2\,J_{5/4}(\frac{e^{2\,x_{{\rm m}}}\,{\nu }^2}{2\,\mu \,\xi })\,
       \left( 4\,x_{{\rm m}} - \log (\frac{8\,{\mu }^2\,{\xi }^2}{{\nu }^4}) \right)  \biggr]}{4\,\xi },
\end{eqnarray}
have been obtained by continuous matching of the solutions in $x_{\rm m}$.
In more explicit terms 
\begin{eqnarray}
A &=& -\pi~ e^{1/8}~ 2^{-5/8}~ \biggl[ 2 Y_{1/4}( \sqrt{2 e}) +\sqrt{2 e} Y_{5/4}(\sqrt{2 e})\biggr] \nu^{-1/2} \mu^{5/4} \xi^{1/4}
\nonumber\\
&=& -1.70146 \times 
\nu^{-1/2} \mu^{5/4} \xi^{1/4},
\nonumber\\
B &=&  \pi ~e^{1/8}~ 2^{-5/8}~ \biggl[ 2 J_{1/4}( \sqrt{2 e}) +\sqrt{2 e} J_{5/4}(\sqrt{2 e})\biggr] \nu^{-1/2} \mu^{5/4} \xi^{1/4} 
\nonumber\\
&=& 3.98472 \times 
\nu^{-1/2} \mu^{5/4} \xi^{1/4}.
\end{eqnarray}
According to Fig. \ref{FIG2}, the analytical approximation, based on the 
continuity of the solutions of Eqs. (\ref{solt}) and (\ref{solosc})
compares very well with the numerical calculation.

Using the results for the evolution of $\epsilon(x)$ the amount of the 
fluctuations produced during the oscillating phase can be estimated.
In Figs. \ref{FIG3} and \ref{FIG4}, the numerical evolution  
for $ \Phi_{k}$ and ${\cal R}_{k}$ are reported. After a flat 
plateau corresponding to the phase where 
\begin{figure}
\centerline{\epsfxsize = 11cm  \epsffile{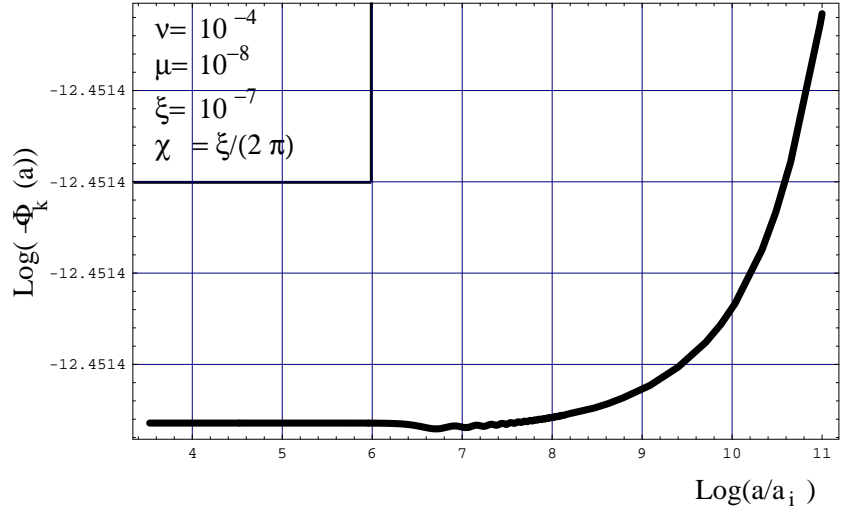}}
\vskip 3mm
\caption[a]{The numerical  evolution of $\Phi_{k}$ illustrated. Notice that the constant value of $\Phi_{k}$ is correctly reproduced by 
 the analytical estimate given in Eq. (\ref{asym1}). }
\label{FIG3} 
\end{figure}
$r(x)$ is constant, the adiabatic fluctuations grow. However, the final value of the adiabatic fluctuations
 (computed, for instance, at 
the moment when $\epsilon$ decays ) will always be very small. 
\begin{figure}
\centerline{\epsfxsize = 11cm  \epsffile{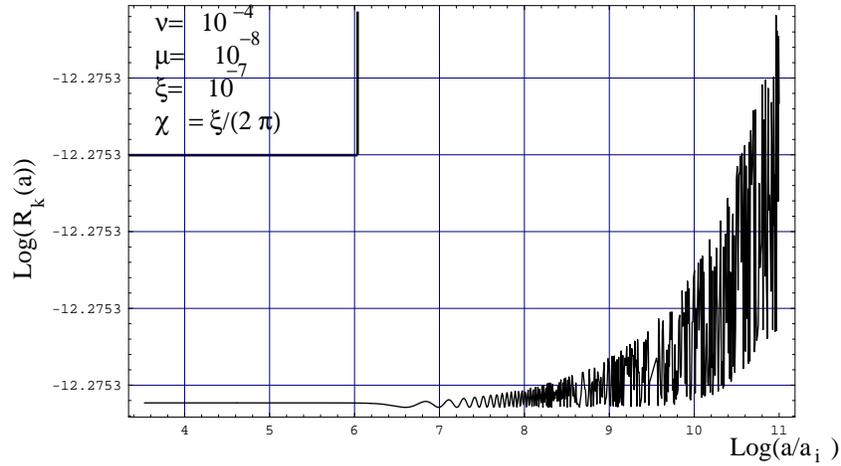}}
\vskip 3mm
\caption[a]{The numerical evolution evolution of ${\cal R}_{k}$ is reported for the same set of parameters discussed in Fig.\ref{FIG4}.}
\label{FIG4} 
\end{figure}
In the integrations reported in Figs. \ref{FIG3} and \ref{FIG4} initial 
conditions are set according to Eqs. (\ref{incon1}).
In order to understand the smallness of the final value of the adiabatic fluctuations, 
recall that the  asymptotic solution of the perturbation equations during the tracking can
be written as
\begin{equation}
 \Phi_{k}^{({\rm m})} = - \frac{2}{9} \mu \tilde{\chi}_{k}^{(i)},~~~~~~~ \chi_{k}^{({\rm m})} =- \frac{4}{9} \mu^2 \tilde{\chi}_{k}^{({\rm i})},
\label{asym1}
\end{equation}
implying  that 
\begin{equation}
{\cal R}_{k}^{({\rm m})} = - \frac{3}{2} \Phi_{k} - \frac{1}{2} \frac{d \Phi_{k}}{d x} =\frac{1}{3} \mu \tilde{\chi}_{k}^{(i)}.
\label{asym2}
\end{equation}
The values obtained in Eqs. (\ref{asym1}) and (\ref{asym2})  are in excellent 
agreement with the amplitude of the flat plateau occurring prior to $x_{\rm m}$ (which 
is about $6.5$ for the parameters chosen in Figs.  \ref{FIG3} and \ref{FIG4}). Different choice of parameters
lead to the same qualitative features in the evolution of the fluctuations.

During the oscillating regime 
\begin{equation}
\frac{ d {\cal R}_{k}}{d x} \simeq \frac{\partial W}{\partial \epsilon} \frac{\chi_{k}}{\rho_{\rm r}},~~~~~~~ r(x)= 4 \mu^2 e^{(x - x_{\rm m})}, ~~~~~~~~ x> x_{\rm m}.
\label{evolr}
\end{equation}
These solutions hold down to the moment of decay occurring at a typical 
curvature scale 
\begin{equation}
H_{\rm d} \sim \biggl(\frac{M^2}{\Lambda}\biggr)^3 M_{\rm P}^{-2},
\label{hd}
\end{equation}
where it has been taken into account that the effective mass is, during the 
oscillating phase $ M^2/\Lambda$.
Integrating the evolution equation for ${\cal R}_{k}$ it can  indeed be
obtained using Eqs. (\ref{asym1})--(\ref{evolr}) 
\begin{equation}
{\cal R}_{k}^{({\rm d})}= {\cal R}_{k}^{({\rm m})} +2 \frac{r_{\rm d}}{\mu} \chi_{k}^{({\rm m})}\simeq \frac{\mu}{3} \biggl( 1 - 
\frac{8}{3} r_{\rm d} \biggr) \tilde{\chi}_{k}^{({\rm i})},
\label{Rd}
\end{equation}
where $r_{\rm d} = r(x_{\rm d})$ and $x_{\rm d}$ is determined through Eq. (\ref{hd}). 
Since $\mu\ll 1$ (to ensure that $\epsilon$ does not dominate 
already during the tracking phase) and since $r_{\rm d} \leq 1$, Eq. (\ref{Rd}) implies that the final value ${\cal R}_{k}^{({\rm d})}$ will be much 
smaller than $\tilde{\chi}_{k}^{({\rm i})}\sim \xi/(2\pi)$.

\renewcommand{\theequation}{4.\arabic{equation}}
\setcounter{equation}{0}
\section{Minimal Inflationary scale} 
In the standard curvaton scenario the energy density of the curvaton
increases with time with respect to the energy density of the 
radiation background. From this aspect of the theoretical construction, a number of constraints 
can be derived; these include an important aspect of the 
inflationary dynamics occurring prior to the curvaton oscillations, namely 
the minimal curvature scale  at the end of inflation compatible with the curvaton 
idea. It has been shown, in the previous sections, that 
to have a phase of tracking curvaton, unfortunately, does not help. 
In the present Section the bounds on the inflationary curvature scale 
will be reviewed, with particular attention to the case where 
the post-inflationary phase is not suddenly dominated by radiation 
like in the case of quintessential inflation \cite{pv,giovannini2}.

\subsection{The standard argument}
First the standard argument will be reviewed (see \cite{lyth4} for a particularly 
lucid approach to this problem). Suppose, for simplicity, that the curvaton field $\epsilon$ 
 has a massive potential 
and that its evolution, after the end of inflation, occurs during a radiation dominated 
stage of expansion.
The field $\epsilon$ starts oscillating at a typical scale 
$H_{\rm m}\sim m$ and the ratio between the curvaton energy density 
and the energy density of the radiation background 
is, roughly
\begin{equation}
r(t) \simeq \biggl(\frac{\epsilon_{\rm i}}{M_{\rm P}}\biggr)^2 \biggl(\frac{a}{a_{\rm m}}\biggr)~~~~~H<H_{\rm m}.
\label{r1}
\end{equation}
When $\epsilon$ decays the ratio $r$ gets frozen to its value at decay, i.e.  $r(t) \simeq 
r(t_{\rm d}) = r_{{\rm d}}$ for $t > t_{\rm d}$. Equation (\ref{r1}) then implies  
\begin{equation}
m = \frac{\epsilon_{\rm i}^2}{ r_{\rm d} M_{\rm P}}. 
\label{mrd}
\end{equation}
The energy density 
of the background fluid just before decay has to be larger
 than the energy density of the decay products, i.e. 
$\rho_{\rm r}(t_{\rm d}) \geq T_{\rm d}^4 $.
Since 
\begin{equation}
\rho_{\rm r}(t_{\rm d}) \simeq m^2 M_{\rm P}^2 \biggl(\frac{a_{\rm m}}{a_{\rm d}}\biggr)^4,
\end{equation}
the mentioned condition implies 
\begin{equation}
\frac{m}{T_{\rm d}} \sqrt{\frac{m}{M_{\rm P}}} > 1,
\end{equation}
which can also be written, using Eq. (\ref{mrd}), as 
\begin{equation}
\biggl( \frac{\epsilon_{\rm i}}{M_{\rm P}}\biggr)^3  \geq r_{\rm d}^{3/2} \biggl(\frac{T_{\rm d}}{M_{\rm P}}\biggr).
\label{epscond1}
\end{equation}
Equation (\ref{epscond1}) has to be compared  with the restrictions 
coming from the amplitude of the adiabatic perturbations, which should 
be consistent with observations. If $\epsilon$ decays before becoming dominant the curvature 
perturbations at the time of decay  are
\begin{equation}
{\cal R}(t_{\rm d}) \simeq \biggl.\frac{1}{\rho_{\rm r}} \frac{\partial W}{\partial \epsilon} 
\chi \biggr|_{t_{\rm d}}\simeq r_{\rm d} \frac{ \chi^{{\rm i}}_{k}}{\epsilon_{\rm i}}.
\end{equation}
Recalling that $\chi^{\rm i}_{k} \sim H_{\rm i}/(2 \pi)$  the power spectrum of curvature perturbations 
\begin{equation}
{\cal P}^{1/2} \simeq \frac{ r_{\rm d} H_{\rm i}}{ 4 \pi \epsilon_{\rm i}} \simeq 5 \times 10^{-5}
\label{pscp}
\end{equation}
implies, using Eq. (\ref{epscond1}) together with Eq. (\ref{mrd}), 
\begin{equation}
\biggl( \frac{H_{\rm i}}{M_{\rm P}}\biggr) 
\geq 10^{-4} \times r_{\rm d}^{-1/2} \biggl(\frac{T_{\rm d}}{M_{\rm P}}\biggr)^{1/3}.
\label{bound1}
\end{equation}
Recalling now that $ r_{\rm d} \leq 1$ and $m < H_{\rm i}$, the above inequality 
implies that, at most, $H_{\rm i} \geq 10^{-12} ~M_{\rm P}$ if $T_{\rm d} \sim 1 ~{\rm MeV}$ is selected.
This estimates is, in a sense, general since the specific relation between $T_{\rm d}$ and $H_{\rm d}$ is not fixed.
The bound (\ref{bound1}) can be even more constraining, for certain 
regions of parameter space, if the condition $ T_{\rm d} \geq \sqrt{H_{\rm d} ~M_{\rm P}}$ is imposed with $H_{\rm d} \simeq m^3/M_{\rm P}^2$.
In this case, Eq. (\ref{bound1}) implies $H_{\rm i}^2 \geq 10^{-8} m ~M_{\rm P}$,  which is 
more constraining than the previous bound for sufficiently large values of the mass, i.e. $ m \geq 10^{-4} H_{\rm i}$.
Thus, in the present context, the inflationary curvature scale is bound to be 
in the interval $10^{-12} M_{\rm P} \leq H_{\rm i} \leq 10^{-6} M_{\rm P} $.

\subsection{Different post-inflationary histories}
Consider now the case where the inflationary epoch is not immediately 
followed by radiation. Different models of this kind may be constructed. 
For instance, if the inflaton field is identified with the quintessence field, a long 
kinetic phase occurred prior to the usual radiation-dominated stage of expansion \cite{pv}.
The evolution of a massive curvaton field in quintessential inflationary models 
has been recently studied \cite{giovannini2} in the simplest scenario where the curvaton field
is decoupled from the quintessence field and it is minimally coupled to the metric.
In order to be specific, suppose that, as in \cite{pv}, the inflaton 
potential, $V(\varphi)$ is chosen to be 
a typical power law during inflation and an {\em inverse} power  during 
the quintessential regime:
\begin{eqnarray}
&&V(\varphi) = \lambda ( \varphi^4 + M^4) ,~~~~ \varphi < 0,
\nonumber\\
&& V(\varphi) = \frac{\lambda M^8}{\varphi^4 + M^4}, ~~~~ \varphi \geq 0,
\label{potph}
\end{eqnarray}
where $\lambda$ is the inflaton self-coupling and  $M$ is the typical 
scale of quintessential evolution. The potential of the curvaton may be 
taken to be, for simplicity, quadratic. In this model the curvaton 
will evolve, right after the end of inflation, in an environnment 
dominated by the kinetic energy of $\varphi$. The curvaton starts oscillating at $H_{\rm m} \sim m$ and 
becomes 
dominant at a typical curvature scale $H_{\epsilon} \sim m (\epsilon_{\rm i}/M_{\rm P})^2$. 
Due to the different 
evolution of the background geometry, the ratio $r(t)$ will take the form
\begin{equation}
r(t) \simeq m^2 \biggl(\frac{\epsilon_{\rm i}}{M_{\rm P}}\biggr)^2 \biggl(\frac{a}{a_{{\rm m}}}\biggr)^3, ~~~~~~H<H_{\rm m}.
\label{kinback}
\end{equation}
to be compared with Eq. (\ref{r1}) valid in the standard case.
For $t>t_{\rm d}$, $r(t)$ gets frozen to the value $r_{\rm d}$ whose relation to 
$\epsilon_{\rm i}$ is different from the one obtained previously (see Eq. (\ref{mrd})) and valid 
in the case when $\epsilon$ relaxes in a radiation dominated environnment. In fact, from Eq. (\ref{kinback}),
\begin{equation}
m \simeq \frac{\epsilon_{\rm i}}{\sqrt{r_{\rm d}}}.
\label{mrdkin}
\end{equation}
From the  requirement
\begin{equation}
\rho_{\rm k}(t_{\rm d}) \simeq m^2 M_{\rm P}^2 \biggl(\frac{a_{\rm m}}{a_{\rm d}}\biggr)^6 \geq T_{\rm d}^4,
\end{equation}
it can be inferred, using (\ref{mrdkin}), that
\begin{equation}
\biggl(\frac{\epsilon_{\rm i}}{M_{\rm P}} \biggr)^{3/2} \geq r_{\rm d}^{3/4} \biggl(\frac{T_{\rm d}}{M_{\rm P}}\biggr) 
\end{equation}
Following the analysis reported in \cite{giovannini2}, the amount of produced fluctuations can be computed.
The spatial curvature perturbation can be written, in this model, as 
\begin{equation}
{\cal R} = - \frac{H}{\dot{\varphi}^2 + \dot{\epsilon}^2} 
\bigl[ \dot{\varphi} v_{\varphi}  +\dot{\epsilon} v_{\epsilon} \bigr],
\label{Rdef1}
\end{equation}
where 
\begin{eqnarray}
&& v_{\varphi} = \chi_{\varphi} + \frac{\dot{\varphi}}{H} \Phi,
\label{defvph}\\
&& v_{\psi} = \chi_{\epsilon} + \frac{\dot{\epsilon}}{H} \Phi,
\label{defvps}
\end{eqnarray}
are, respectively,  the canonically normalized fluctuations of $\varphi$ and $\epsilon$. As discussed in \cite{giovannini2} the relevant evolution equations 
can be written as 
\begin{eqnarray}
&&\ddot{v}_{\varphi} + 3 H \dot{v}_{\varphi} - \frac{1}{a^2} \nabla^2 v_{\varphi} +
\biggl[ \frac{\partial^2 V}{\partial\varphi^2} - \frac{1}{M_{\rm P}^2 a^3}\frac{\partial}{\partial t} 
\biggl( \frac{a^3}{H} \dot{\varphi}^2\biggr)\biggr] v_{\varphi} 
- \frac{1}{M_{\rm P}^2 a^3}\frac{\partial}{\partial t} 
\biggl( \frac{a^3}{H} \dot{\varphi} \dot{\epsilon}\biggr) v_{\epsilon}=0,
\label{vph}\\
&&\ddot{v}_{\epsilon} + 3 H \dot{v}_{\epsilon} 
- \frac{1}{a^2} \nabla^2 v_{\epsilon} +
\biggl[ \frac{\partial^2 W}{\partial\epsilon^2} 
- \frac{1}{M_{\rm P}^2 a^3}\frac{\partial}{\partial t} 
\biggl( \frac{a^3}{H} \dot{\epsilon}^2\biggr)\biggr] v_{\epsilon} 
- \frac{1}{M_{\rm P}^2 a^3}\frac{\partial}{\partial t} 
\biggl( \frac{a^3}{H} \dot{\varphi} \dot{\epsilon}\biggr) v_{\varphi}=0.
\label{vps}
\end{eqnarray}
Solving Eqs. (\ref{vph}) and (\ref{vps}) with the appropriate initial condition, and using that 
$a^3 \dot{\varphi}^2/H $ is constant during the 
kinetic phase, Eq. (\ref{Rdef1}) can be written as  \cite{giovannini2}
\begin{equation}
{\cal R}(t) \simeq \frac{H}{\rho_{\rm k}} \dot{\epsilon} v_{\epsilon} 
\simeq \frac{\chi_{\epsilon}}{\rho_{\rm k}} \frac{\partial W}{\partial\epsilon}\simeq
 r_{\rm d}\biggl(\frac{\chi^{({\rm i})}_{k}}{\epsilon_{\rm i}}\biggr).
\end{equation}
Recalling   that $\chi^{({\rm i})}_{k} \sim H_{\rm i}/(2 \pi)$, the observed 
value of the power spectrum, i.e.  
$ {\cal P}^{1/2}_{{\cal R}} \sim 5 \times 10^{-5}$, implies 
\begin{equation}
\biggl(\frac{H_{\rm i}}{M_{\rm P}} \biggr) \geq 10^{-4} \biggl(\frac{T_{\rm d}}{M_{\rm P}}\biggr)^{2/3}. 
\label{bound2}
\end{equation}
The same approximations discussed in the standard case will now be applied to the case 
of quintessential inflation.
Suppose that  $T_{\rm d} \sim 1 {\rm MeV}$ (i.e. the minimal value compatible with nucleosynthesis considerations). Thus, from Eq. (\ref{bound2}) 
$H_{\rm i} \geq 10^{-19} M_{\rm P}$. This estimate has to be compared  with $H_{\rm i} \geq 10^{-12} ~M_{\rm P}$ which was 
obtained (see Eq. (\ref{bound1})) in the case when $\epsilon$ relaxes to its minimum in a radiation dominated environnment.
Suppose now that the decay of $\epsilon$ is purely gravitational, i.e. $\Gamma \sim m^3/M_{\rm P}$.
If this is the case, as previously argued,  one could also require that $T_{\rm d} \geq \sqrt{H_{\rm d} M_{\rm P}}$.  
This requirement implies that for masses $m > 10 {\rm TeV}$, 
$H_{\rm i} > 10^{-4} m$. 

\renewcommand{\theequation}{5.\arabic{equation}}
\setcounter{equation}{0}
\section{Concluding remarks} 
In the present investigation the possibility that the ratio between the curvaton energy density 
and that of the dominant component of the background sources is constant during 
a significant part in the evolution of the Universe. The 
various estimates of this preliminary analysis seem to suggest that if $r'\simeq 0$ 
for a sufficiently long time the  obtained curvature perturbations 
present prior to curvaton decay are much smaller than the value required by observations.
The possibility of having $r'\simeq 0$ down to sufficiently low curvature scales would, on the other hand,
be interesting  to relax the bounds on the minimal 
inflationary curvature scale, which, in the standard scenario (where $r'>0$), is 
roughly $H_{\rm i} > 10^{-12} M_{\rm P}$ for the most optimistic 
set of parameters. If this bound could be evaded inflation could take place, within 
the framework discussed in the present paper, also at much smaller curvature scales. 
In this sense, the presence of a phase $r'=0$ does not alleviate the problem. Furthermore, if 
$r'=0$ for a  sufficiently long time, the final amount of curvature perturbations gets drastically reduced.

In a related perspective the case of low-scale quintessential inflation has been examined. 
In this case, $r'>0$ and the background geometry is kinetically dominated down to the 
moment of curvaton oscillations. The same argument, leading to the standard 
bound on the minimal inflationary  curvature scale shows, in this case, that 
the bounds are a bit relaxed and curvature scales $H_{\rm i} > 10^{-19} M_{\rm P}$ become plausible.


\begin{thebibliography}{99}

\bibitem{lyth1} D. Lyth and D. Wands, Phys. Lett. B{\bf 524}, 5 (2002). 

\bibitem{moroi} T. Moroi and Takahashi, Phys. Rev. D {\bf 66}, 063501 (2002).

\bibitem{enqvist} K. Enqvist and M. Sloth, Nucl.Phys.B {\bf 626}, 395 (2002).

\bibitem{sloth} M. Sloth, Nucl.Phys.B {\bf 656}, 239 (2003).

\bibitem{giovannini1} V. Bozza {\it et al}., Phys.\ Lett.\ B {\bf 543}, 14 (2002);
Phys.Rev.D {\bf 67}, 063514 (2003). 

\bibitem{mollerach} S.~Mollerach,
Phys.\ Rev.\ D {\bf 42}, 313 (1990).

\bibitem{bartolo} N.~Bartolo and A.~R.~Liddle,
Phys.\ Rev.\ D {\bf 65}, 121301 (2002).

\bibitem{lyth2} 
D.~H.~Lyth, C.~Ungarelli and D.~Wands,
Phys.\ Rev.\ D {\bf 67}, 023503 (2003)

\bibitem{giovannini2}
M.~Giovannini,
Phys.\ Rev.\ D {\bf 67}, 123512 (2003).

\bibitem{basterogil} M.~Bastero-Gil, V.~Di Clemente and S.~F.~King,
Phys.\ Rev.\ D {\bf 67}, 083504 (2003); Phys.\ Rev.\ D {\bf 67}, 103516 (2003).

\bibitem{enqvist1} 
K.~Enqvist, A.~Jokinen, S.~Kasuya and A.~Mazumdar, hep-ph/0303165; 

\bibitem{enqvist2}K.~Enqvist, S.~Kasuya, A.~Mazumdar,  Phys. Rev. Lett. {\bf 90} 091302 (2003). 

\bibitem{lyth3} K. Dimopoulos {\it et al.}, JHEP {\bf 0305}, 057 (2003); 
hep-ph/0308015. 

\bibitem{lyth4} D. Lyth, hep-th/0308110.

\bibitem{dimopoulos} K. Dimopoulos, astro-ph/0212264.

\bibitem{fujii} M. Fujii and T. Yanagida, Phys. Rev. D {\bf 66} 123515 (2002).

\bibitem{pv} P.J.E. Peebles, A. Vilenkin Phys. Rev. D {\bf 59}, 063505 (1999). 

\bibitem{giovannini3} M.~Giovannini, Phys.\ Rev.\ D {\bf 60}, 123511 (1999); 
Class.\ Quant.\ Grav.\  {\bf 16}, 2905 (1999).

\bibitem{barreiro} 
T.~Barreiro, E.~J.~Copeland and N.~J.~Nunes,
Phys.\ Rev.\ D {\bf 61}, 127301 (2000).

\bibitem{albrecht}
A.~Albrecht and C.~Skordis,
Phys.\ Rev.\ Lett.\  {\bf 84}, 2076 (2000).

\bibitem{sahni} V. Sahni and L. Wang, Phys. Rev. D {\bf 62}, 103517 (2000).

\bibitem{rp1}
P.~J.~Peebles and B.~Ratra,
Rev.\ Mod.\ Phys.\  {\bf 75}, 559 (2003).
\end{thebibliography}
\end{document}